\documentclass{article}
\usepackage{arxiv}
\usepackage[utf8]{inputenc} 
\usepackage[T1]{fontenc}    
\usepackage{hyperref}       
\usepackage{url}            
\usepackage{booktabs}       
\usepackage{amsfonts}       
\usepackage{nicefrac}       
\usepackage{microtype}      
\usepackage{lipsum}		
\usepackage{graphicx}
\usepackage{natbib}
\usepackage{doi}
\usepackage[utf8]{inputenc} 
\usepackage[T1]{fontenc} 
\usepackage{hyperref} 
\usepackage{url} 
\usepackage{booktabs} 
\usepackage{amsfonts} 
\usepackage{nicefrac} 
\usepackage{microtype} 
\usepackage{lipsum}
\usepackage{todonotes}
\usepackage{subcaption}
\usepackage[utf8]{inputenc}
\usepackage{enumitem}
\usepackage{cite}
\usepackage[]{microtype}

\usepackage{scrextend}
\setlist{nosep}
\renewcommand{\paragraph}[1]{\noindent {\bf #1}}

\usepackage{bbding}

\title{Game and Simulation Design for Studying Pedestrian-Automated Vehicle Interactions}

\author{\href{https://orcid.org/0000-0003-2046-4622}{\includegraphics[scale=0.06]{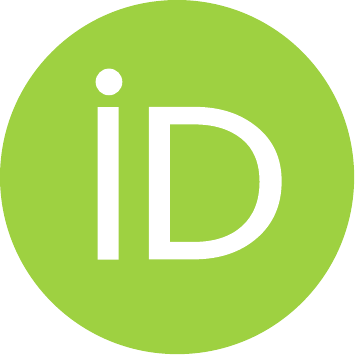}\hspace{1mm}Georgios Pappas} \\
	Department of Electrical and Computer Engineering\\
	Michigan State University\\
	East Lansing, MI, 48824 \\ \\
	Department of Electrical and Computer Engineering\\
	National Technical University of Athens\\
	Athens, Greece\\ \\ 
	Lab of Educational Material and Methodology\\
	Open University of Cyprus\\
	Nicosia, Cyprus\\
	\texttt{geopap1987@hotmail.com} \\
	\And
	\href{https://orcid.org/0000-0002-5540-7401}{\includegraphics[scale=0.06]{orcid.pdf}\hspace{1mm}Joshua E. Siegel} \\
	Department of Computer Science and Engineering\\
	Michigan State University\\
	East Lansing, MI, 48824 \\
	\texttt{jsiegel@msu.edu} \\
	\And
	\href{https://orcid.org/0000-0002-6257-2253}{\includegraphics[scale=0.06]{orcid.pdf}\hspace{1mm}Jacob Rutkowski} \\
	Department of Computer Science and Engineering\\
	Michigan State University\\
	East Lansing, MI, 48824 \\
	\texttt{rutkow72@msu.edu} \\
	\And
	{Andrea Schaaf} \\
	Department of Communication\\
	Michigan State University\\
	East Lansing, MI, 48824 \\
	\texttt{schaafan@msu.edu} \\
}

\hypersetup{
pdftitle={Game and Simulation Design for Studying Pedestrian-Automated Vehicle Interactions},
pdfsubject={},
pdfauthor={Georgios Pappas, Joshua E.~Siegel, Jacob Rutkowski, Andrea Schaaf},
pdfkeywords={Gamification, Simulation, Self-Driving Vehicles, Pedestrian Interaction},
}

\begin{document}
\maketitle

\begin{abstract}
The present cross-disciplinary research explores pedestrian-autonomous vehicle interactions in a safe, virtual environment. We first present contemporary tools in the field and then propose the design and development of a new application that facilitates pedestrian point of view research. We conduct a three-step user experience experiment where participants answer questions before and after using the application in various scenarios. Behavioral results in virtuality, especially when there were consequences, tend to simulate real life sufficiently well to make design choices, and we received valuable insights into human/vehicle interaction. Our tool seemed to start raising participant awareness of autonomous vehicles and their capabilities and limitations, which is an important step in overcoming public distrust of AVs. Further, studying how users respect or take advantage of AVs may help inform future operating mode indicator design as well as algorithm biases that might support socially-optimal AV operation.  
\end{abstract}




\section{Introduction}
Due to the advancements of modern technologies automated vehicles (AVs) are an obvious next step in the evolution of the transportation. AVs have the potential to lead positive societal change, including reduced carbon emissions, travel costs, and transit times~\citep{Tirachini2020,Piao2016,Deb2018,Habibovic2018}. Beyond these impacts, AVs bring about the potential for enhanced safety relative to human driving, as human error is a leading cause of road accidents as a result of driver distraction and other factors~\citep{Piao2016}. Despite AV's potential benefits, there seems to be a distrust about their safety~\citep{Deb2018, Kassens-Noor2020}. In particular, AV-pedestrian interaction poses unknowns, in part due to the complexity of capturing real-world data from pedestrians without undue risk. In order to safely conduct this research, there is an opportunity to instead study interactions in (realistic) simulated virtual environments. 

Studies have demonstrated simulations' ability to generate realistic data, including for AV-pedestrian interactions~\citep{Deb2017}. Although  there are some great efforts presented in academia, we also noticed the lack of public availability of such solutions ~\citep{Deb2017}. So, this is an important factor for us since our intention is to release our proposed solution as a free-to-use solution useful even to non-technical researchers. On the other hand some other applications are available for public use, such as AirSim and Carla, though those applications are from the perspective of the automated vehicle, while our application is taking a more pedestrian-centered approach. This allows our application to build on prior art while also setting things from a new perspective that will be necessary in automated vehicle research.

Complementing prior art, our team has developed a first-person simulator that allows participants to control a human avatar within an urban environment to allow individuals interact with virtual automated and virtual human-operated vehicles as they attempt to cross a street. A configurable application allows researchers to effectively capture representative interactions and can also be used as a way to train participants on AV safety while building trust and awareness, easing the safe and efficient adoption of driverless technologies.

The present manuscript describes the process by which our simulator was developed, tested, and validated. Included is a summary of prior art in simulation related to AV's, and we provide comparison of our simulator to those existing in order to identify opportunities. We then discuss goals and applications of our simulator, and experimental results examining human interaction with our tool and resulting iterative design changes. We close by exploring how our tool can be applied to advance the efforts of the AV industry, and by describing planned next steps in using the simulator to capture data for informing AV design and human interaction.


 \section{Motivation}
 The Pedestrian Tool is an easy-to-use application that allows researchers to gather data and better understand how the average pedestrian will interact with automated vehicles. One of the main ideas that our team focused on was accessibility, as we wanted to make sure that the tool was quite user friendly. This will allow users to better understand interaction patterns between humans and automated vehicles through data collection using the tool. Although there have been other tools that are similar to what we have created (see subsection ~\ref{Similar Applications}), these simulations struggle with accessibility issues. Some simulations are private and therefore lack public support and others can be used only by those with an advanced coding skills. Further to this, other tools are mostly vehicle-centered where our work will try to approach the whole problem from the perspective of a pedestrian. So, in short, we have identified the need of an easy-to-use, plug-and-play and pedestrian-centered gamified research tool for the field. Thus, our application will try to bridge this gap offering the community a tool that can be used by professionals in multiple domains.
 
 This tool will also become important due to approaching reality that automated vehicles are beginning to be introduced into the general public, and with this a lot of questions have been raised about their interactions with humans. There is a general distrust surrounding automated vehicles due to a lack of understanding of how they work or what they will do, so this tool will allow researchers to better understand how to combat this distrust while also allowing the general public to interact with automated vehicles in a safe and harmless environment. The stigma labeling automated vehicles as 'unpredictable machines' could create additional challenges against their involvement in society and potential commercial failure, and we believe that their benefits far outweigh their drawbacks. So, one of the main goals of the tool is to raise awareness and improve the public perception towards automated vehicles and their safety.

\section{Prior Art}
AV's have the potential to benefit the efficiency and environmental friendliness of our world ~\citep{Piao2016,Deb2018,Habibovic2018}. As a result, most individuals surveyed would consider using automated vehicles if and when they become available for general public use~\citep{Piao2016, Kassens-Noor2020}. However, some people do not trust AV's, citing concerns about privacy, safety, and uncertainties with regards to algorithm performance. ~\citep{Piao2016,Shahrdar2019,Ackermann2019,Keeling2019}. Reig et al. concluded that the general public who do not support AVs may not understand AV capabilities and lack education on the topic~\citep{Reig2018}, suggesting that education may drive support. Even this, however, may not be sufficient to comfort individuals concerned about uncertain algorithm performance, such as how vehicles will perform in the absence of sensor data or external communications~\citep{Li2018,Chen2017}. Similarly, a lack of AV data provided to humans may in fact lead to reduced AV performance. Studies are therefore exploring the best ways for autonomous vehicles to communicate information with human drivers and pedestrians in real time, better allowing humans to predict the algorithm's behavior and plan accordingly~ \citep{Habibovic2018,Fisac2019}.

Pedestrian/AV safety is an ongoing research concern. Exploration in the area of pedestrian safety has historically been limited by the constant inability to put real human subjects into dangerous situations~\citep{Deb2017}. Recently, the use of virtual environments to research pedestrian-automated vehicle interactions has been explored, with Deb et al. concluding that virtual environments can simulate pedestrian actions with sufficient realism to mirror real-world results~\citep{Deb2017}. Jayaraman et al. applied a similar concept, using the Unity Game Engine to develop a simulator that collected information on pedestrian-AV interactions~\citep{Jayaraman2020}, creating ``an accurate long-term pedestrian crosswalk behavior (trajectory) prediction model'' from synthetic data~\citep{Jayaraman2020}.

Many of these tools and applications are designed using popular game engines. The Unity Game Engine is one option that has been used successfully in pedestrian simulation~\citep{Jayaraman2020}. This multi-platform engine allows developers to create simulated environment where they can spawn or destroy of (Game)objects~\citep{Juliani2018,Pappas2021}. The design and development  process of a tool has become quite easier since a great variety of 3D objects can be acquired through Unity's extensive asset store that eases access to CAD or even scriptable physics models\citep{Juliani2018,Nguyen2017}. This flexibility allows researchers to develop simulations, including gamified solutions, iteratively in order to respond to user feedback, particularly at early stages\citep{Morschheuser2018}.

Another important factor in doing virtual research is determining the ideal mode of interaction (2D, 3D, or some form of XR including VR, AR, or MR). Virtual Reality (VR) allows for participants to feel as though their actions have real consequences~\citep{Nguyen2017,Fernandez2017}. However, working with virtual reality can cause ``virtual reality sickness'' as a result of prolonged time spent in virtual environments~\citep{Deb2017}, suggesting breaks may need to be part of game design. Despite this, VR has been used successfully in AV studies such as Shahrdar et. al's research. This experiment has participants immerse themselves in a VR simulation and test their trust of virtual AVs after that vehicle made a mistake~\citep{Shahrdar2019}. The study found that most participants were able to regain trust in an AV shortly after the vehicle had made an error~\citep{Shahrdar2019}. Extended Reality (XR) and Mixed Reality (MR) can also be used in future experiments involving AVs, allowing for more information to be communicated to passengers and pedestrians as well as helping build trust between humans and AVs ~\citep{Riegler2020}.


\subsection{Embodiment in Virtual Environments and Video Games} \label{Embodiment}
Virtual Environments and and video games captivate their audiences by allowing individuals access to non-typical experiences. This corresponds with the idea of embodiment, in which a player assumes the role of a virtual character and in so doing subconsciously extends themselves to inherit that character's goals and aspirations, strengths, and limitations~\citep{Gee2008,Moghimi2016}. Embodiment is an important concept in video games as player engagement depends upon how well the player can take control of their virtual character; without sufficient interactivity, games become similar to other forms of media including films or books. In order to have a strong sense of embodiment, one must design a virtual world around the character that helps fully realize the goals of the character while also playing to the character’s strengths~\citep{Gee2008}. This allows the player the ability to use the character's affordances effectively to achieve an end goal.

Embodiment is also important in gamified research, although there is debate about the efficacy of embodiment and its limitations on immersion. Alton raises the question of whether or not one should refer to current virtual character-player relationships as embodiment, because embodiment requires physical reactions and feelings that come as a result of the virtual experience~\citep{Alton2017}. It may be argued whether or not these criteria are met by VR or even video games in general, which leads to some ambiguity in the term of embodiment in virtual gaming. Although this raises concerns about the idea of embodiment in gamified research, embodiment has been proven effective in virtual environments that allow users to take control of an avatar and explore, e.g. a study done by Kiela et al. found that embodiment in video games is a valid first step to developing AI~\citep{Kiela2016}. In this study, humans virtually simulate what a human would and should do in a situation, and then an AI can uncover patterns from the humans, building upon these to refine an algorithm that can be used to simulate what a human would do in the same situation ~\citep{Kiela2016}. This allows algorithms to constantly improve without feedback from human subjects in the virtual world. Another study by Gonzalez-Franco and Peck found that the use of embodied avatars in VR has led to higher levels of user immersion inside simulation~\citep{Gonzalez-Franco2018}. This is important as the more immersed a user is in a simulation, the more realistic the simulation will feel to the user, leading to the creation of better-representative real-world data. 

\subsection{Pedestrian-related research applications and other software} \label{Similar Applications}
To better understand how our simulation compared to other applications of similar purpose, we decided to explore similar applications to appreciate where other automated vehicle simulators succeeded and fell short as well as how we could set our application apart. In order to do this, we selected three applications, with the first two being chosen as they are some of the most popular in the field of automated vehicle simulation, and the third application as it most closely resembles our goal of focusing more on the pedestrian than the vehicle. The first application was Carla Simulator \citep{Dosovitskiy2017}, which was developed by Carla Team. It’s an open-source automated vehicle simulator built with the Unreal Game Engine that works with the OpenDRIVE standard to define the roads and other urban settings. As mentioned above, Carla has grown to be one of the most admired options for automated vehicle simulations over the years and so it has built a growing community around it.

\begin{figure}
 \centering
 \includegraphics[width=0.6\linewidth]{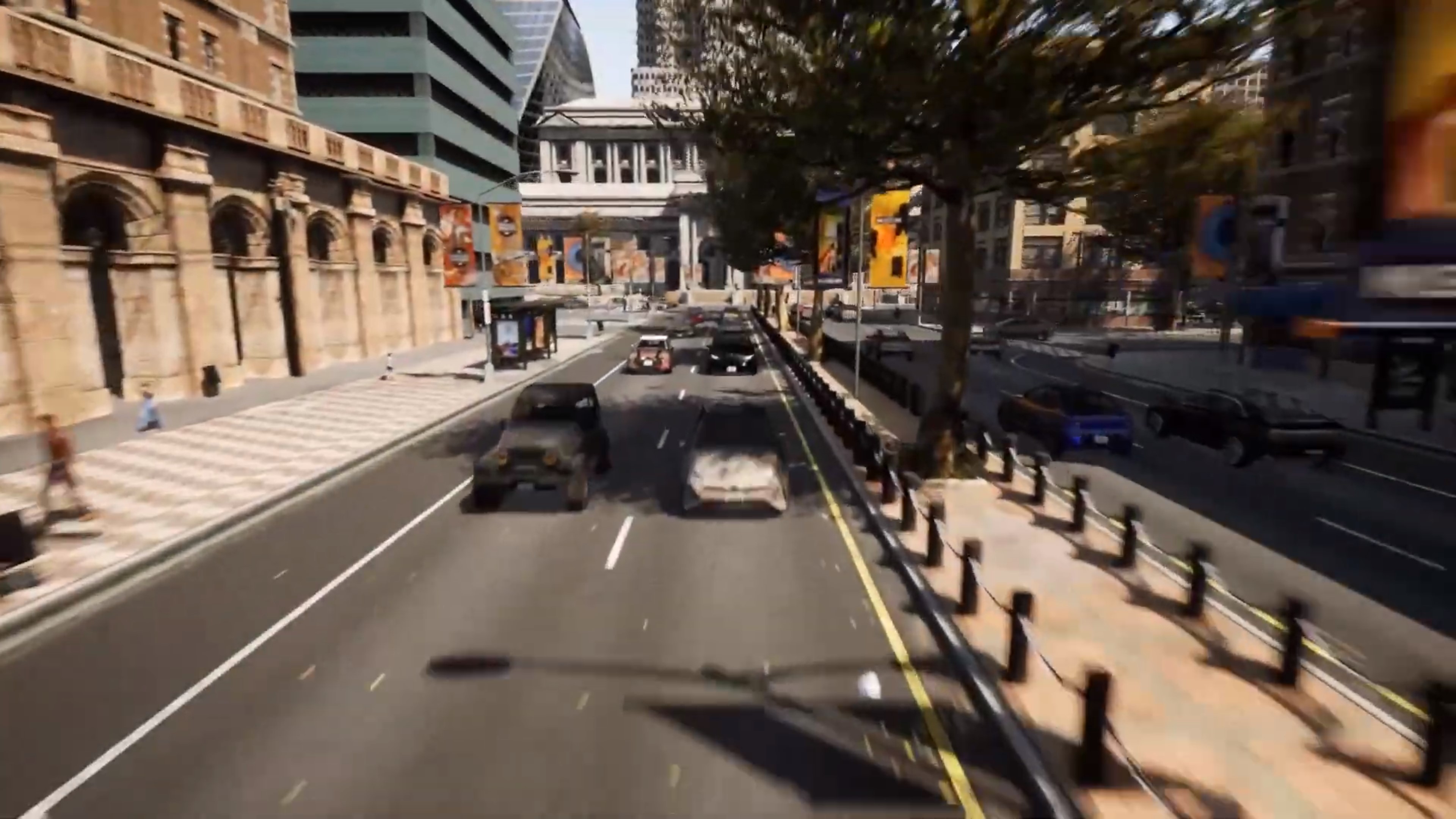} 
 \caption{Screenshot from the Carla Application}
 \label{Carla}
\end{figure}

The second application that we tested was AirSim \citep{Shah2017}, developed by Microsoft Research. It is a computer-based simulator that was built for drone and self-driving car research. Like the Carla application, this simulator is built upon the Unreal Game Engine, but there is also an experimental version based on Unity in the works. This simulator is primarily used for research purposes surrounding AI and experimentation with deep learning, computer vision, and reinforcement language.

\begin{figure}
 \centering
 \includegraphics[width=0.6\linewidth]{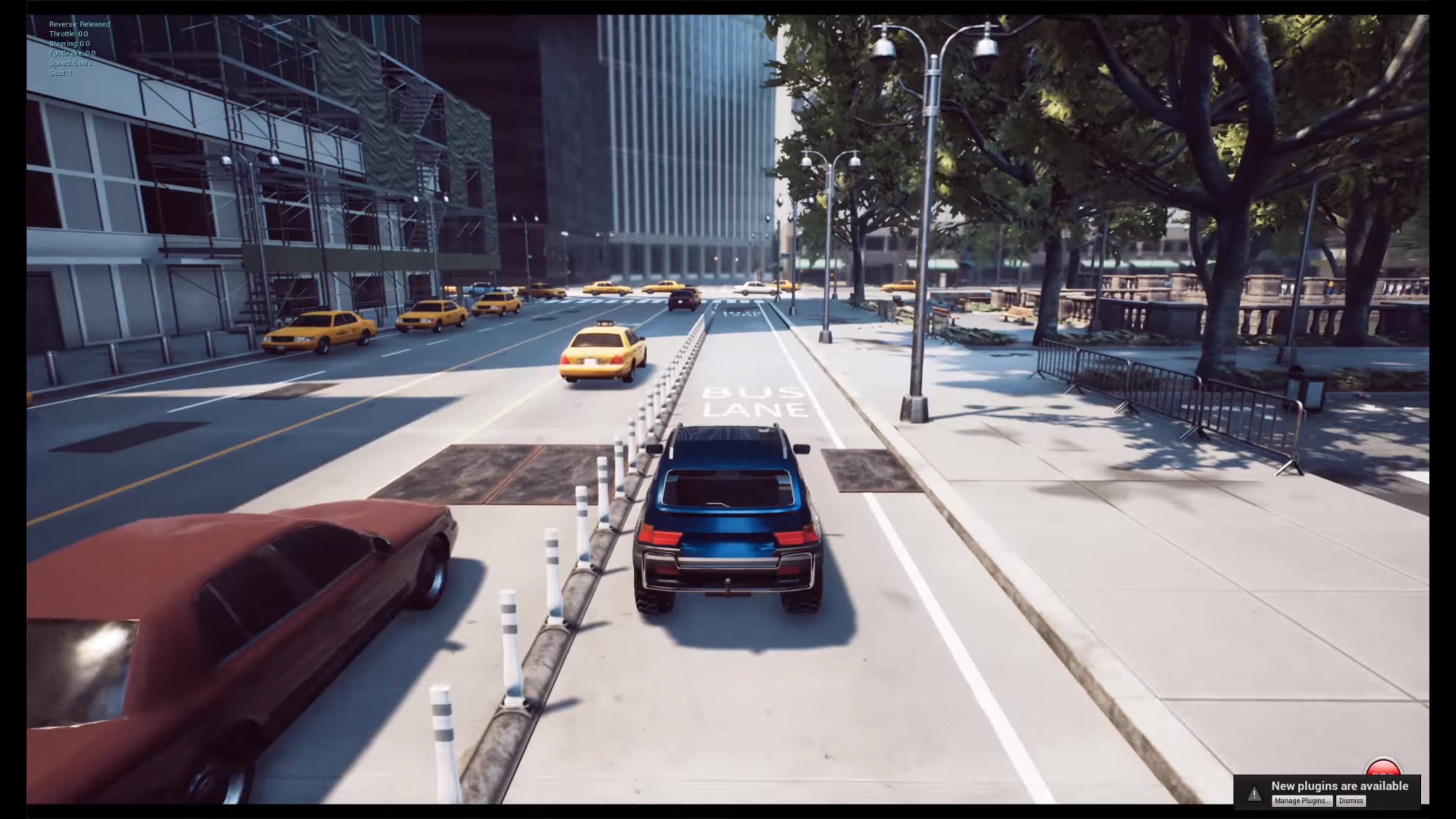} 
 \caption{Screenshot from the AirSim Application}
 \label{Airsim}
\end{figure}

The final application that we looked at when exploring competitive applications was a tool developed by Deb et al. for research at Mississippi State University \citep{Deb2018}\citep{Deb2017}. This simulation is not publicly available, though from the research papers released about it, we can conclude that the simulation focuses on pedestrians crossing an intersection in virtual reality using the HTC Vive headset, which is close to the goal of our team as well. The tool was also developed upon the Unity Game Engine, setting itself apart from the other two applications while also being similar to ours. 

\begin{figure}
 \centering
 \includegraphics[width=0.6\linewidth]{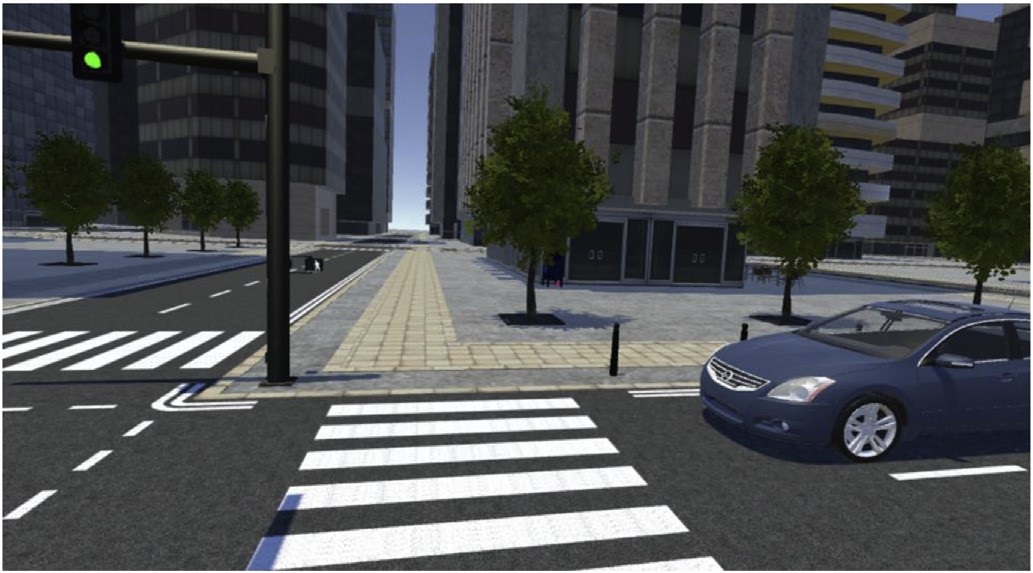} 
 \caption{Screenshot from the Mississippi State Tool}
 \label{MSTool}
\end{figure}

In order to compare these applications, we analyzed each application based on five categories, which were:

\begin{itemize}
    \item \textbf{Availability} – All simulators are targeting the academic community; which simulators are free and/or open-source?
    \item \textbf{Engines and Technical Development} – What game engine are these simulators built on and are there any expandability options?
    \item \textbf{Usability}, Type of Application, and Target Audience – Identify what the target audience is based on what type of prior knowledge the user needs to utilize the simulator and conduct research with the simulator. Regarding types of application, does the application use VR, desktop, or both kinds of implementation?
    \item \textbf{Realistic Graphics Environment} – What is the quality of the 3D environments and models/prefabs used.
    \item \textbf{Pedestrian Point of View} – How is the pedestrian feature implemented and could it be used for training or other types of research?
\end{itemize}

The goal of comparing all of the tools with the aforementioned criteria was to better understand what each of these applications did well so that we could better innovate in our tool while also understanding where these applications did lack so that we could add features to improve the experience and set our application apart from the competition.

\begin{table}[]
    \centering
    \resizebox{0.6\linewidth}{!}{%
    \begin{tabular}{c | c c c}
    \hline
    \hline
     & \textbf{Carla Simulator} & \textbf{AirSim} & \textbf{Mississippi State University Tool} \\
    \hline
    \textbf{Availability} & Free and Open Source & Free and Open Source & Publicly Unavailable\\
    \hline
    \end{tabular}
    }
    \caption{Availability Comparison}
    \label{tab:AvailabilityTable}

\end{table}

The first comparison that we made was the availability of each simulator. As can be seen  at Table \ref{tab:AvailabilityTable}, Carla Simulator and AirSim, the two most popular tools available, are both free and open source. This fact gives these tools a major advantage over the Mississippi State University Tool as they have had years to build a community that supports their applications, and this is why they are considered some of the most dominant applications in the field. Considering the Mississippi State University tool, it has potential based on research results published about it, but it isn’t publicly available and therefore we can’t say if it will become as viable as the other two applications.

\begin{table}[]
    \centering
    \resizebox{0.6\linewidth}{!}{%
    \begin{tabular}{c | c c c}
    \hline
    \hline
     & \textbf{Carla Simulator} & \textbf{AirSim} & \textbf{Mississippi State University Tool} \\
    \hline
    \hline
    \textbf{Unity} & No Compatibility & Experimental Compatibility & Compatible\\
    \hline
    \textbf{Unreal Engine} & Compatible & Compatible & No Compatibility\\
    \hline
    \textbf{Technical} & C++, Python & C++, Python, C\# & No Compatibility\\
    \textbf{Development} & & & \\
        \hline
    \hline
    \end{tabular}
    }
    \caption{Game Engines and Technical Development Comparison}
    \label{tab:GameEngines_Comparison}
\end{table}

The next category compares the applications’ game engine and technical development compatibilities. As shown at Tables \ref{tab:GameEngines_Comparison}, Carla Simulator is built upon the Unreal Engine and therefore is compatible with it, allowing users with knowledge in C++ to work on it. In addition, some knowledge of Python is also required with the deep learning elements of the self-driving vehicle in the virtual environment. AirSim is also built upon the Unreal Engine, and therefore it requires users to have the same C++ and Python knowledge in order to do any technical developments on the application. However, AirSim also has experimental compatibility with the Unity Game Engine, and therefore it increases its community by allowing for technical developments with the C\# programming language and its large community of developers. Finally, the Mississippi State University Tool also taps into this C\# community through its use of Unity Game Engine, but as mentioned before, it is not publicly available and so no technical development can be made to the application.

\begin{table}[]
    \centering
    \resizebox{0.6\linewidth}{!}{%
    \begin{tabular}{c | c c c}
    \hline
    \hline
     & \textbf{Carla Simulator} & \textbf{AirSim} & \textbf{Mississippi State University Tool} \\
    \hline
    \hline
    \textbf{Ease of Use} & Hard to Use & Hard to Use & Easy to use\\
    \hline
    \textbf{Type of Application} & Desktop & Desktop & Virtual Reality\\
    \hline
    \textbf{Target Audience} & Mainly Programmers & Mainly Programmers & Academia Audience\\
    \ & or Engineers & or Engineers &  \\
    \hline
    \hline
    \end{tabular}
    }
    \caption{Usability and Target Audience Comparison}
    \label{tab:Usability_TargetAudience_Comparison}
\end{table}

The usability and target audience category analysis (Table \ref{tab:Usability_TargetAudience_Comparison}) shows that the two main tools, Carla Simulator and AirSim, require strong programming skills in multiple programming languages and are relatively hard to set up, especially for users who are inexperienced. Despite all of their excellent features, this drawback limits the audience of these applications to programmers, engineers, and well-versed automated vehicle enthusiasts. The Mississippi State University Tool succeeds comparatively in this category due to its ease of use and its even easier setup due to it being mostly an export build of a Unity Engine application. This allows the tool to be used by programmers, engineers, enthusiasts, and researchers who may want to conduct research in the simulation relating to automated vehicles and pedestrian interactions. Another major success for the Mississippi State University Tool is that it is a virtual reality application, which allows users to become more immersed in the simulation.

\begin{table}[]
    \centering
    \resizebox{0.6\linewidth}{!}{%
    \begin{tabular}{c | c c c}
    \hline
    \hline
     & \textbf{Carla Simulator} & \textbf{AirSim} & \textbf{Mississippi State University Tool} \\
    \hline
    \textbf{Graphics} & High  & High  &  Average\\
    \hline
    \hline
    \end{tabular}
    }
    \caption{Virtual Environment Visual Fidelity Comparison}
    \label{tab:Graphics_Comparison}
\end{table}

The next category compares how realistic the graphics of each simulation are (Table \ref{tab:Graphics_Comparison}). The first two simulations, Carla Simulator and AirSim, both have multiple scenes and use 3D assets that are of AAA quality, allowing for an incredibly immersive experience within the simulation. The Mississippi State University Tool falls short in this category, despite having average graphics, its graphics don’t compare to those of the first two.

\begin{table}[]
    \centering
    \resizebox{0.6\linewidth}{!}{%
        \begin{tabular}{c | c c c}
        \hline
        \hline
         & \textbf{Carla Simulator} & \textbf{AirSim} & \textbf{Mississippi State University Tool} \\
        \hline
        \textbf{Pedestrian Point of} & No Pedestrian POV & No Pedestrian POV & Pedestrian POV\\
        \textbf{View} & & & \\
        \hline
        \hline
        \end{tabular}
    }
    \caption{Pedestrian Point of View Comparison}
    \label{tab:PedestrianPOV_Comparison}
\end{table}

The final category is the implementation of the pedestrian point of view in the application (Table \ref{tab:PedestrianPOV_Comparison}), which is one of the most important categories for our research. Both Carla Simulator and AirSim lack in this category since their focus is on automated vehicles rather than pedestrian research. Despite being excellent applications, there is no user control of pedestrians as they are more AI-oriented. Both of these applications do, however, allow for customizable scenarios with pedestrians, such as volume and diversity of pedestrian populations. The Mississippi State University Tool focuses on the pedestrian point of view, allowing users to embody the pedestrian in virtual reality..

This competitive analysis allowed us to realize that there is an opportunity to create a new pedestrian-centered application that we will be able to use to conduct training research surrounding automated vehicle awareness. So, in order to allow our application to succeed, we needed to create a human-centered application for training and awareness since making another car-focused simulator would make it hard to compete due to the dominance of the applications Carla Simulator and AirSim. We also intend to make our application publicly available (although not open source) as it would allow our application to have more exposure and feedback for future upgrades. Also, ease of use was a priority for our team as we need our tool to be used not only by those well versed in the technical fields but also by participants in research studies and other scientists with diverse backgrounds. Graphics should also be realistic in order to better simulate and immerse users in the experience. Our application focuses on developing a desktop version with a first-person controller, but we do plan on developing and releasing a virtual reality version in the future as well. Finally, although we developed our application in the Unity Game Engine, some features will be customizable with external files, allowing users and testers to customize scenarios without having to change anything within the source code.

\section{Pedestrian Tool Design and Development}
\subsection{Methods}
The designing process of our tool, was really important. Following the principles of research-creation~\citep{lelievre:hal-02615671}, we needed to create a meaningful~\citep{K.Salen} interactive experience to our users. Thus, we needed to design a serious gamified tool~\citep{Ott2011} that builds awareness and at the same time educates~\citep{D.R.Michael2005} the participants on autonomous vehicles via a realistic simulation environment.

Very significant part of our design methodology was the embodiment element since users will take the role a virtual pedestrian. This role of embodiment is explained in more detail at the Subsection \ref{Embodiment}.  

In terms of development, our application was developed using the Agile methodology and more specifically, the Scrum model~\citep{Jesse2014}. Practically, we developed and conducted parts of our research with stable builds of our tool, working in sprints, while we continued iterating~\citep{Zimmerman2003} until we reached stable builds.

\subsection{Development}
\subsubsection{Scenes and Scenarios} \label{Scenes-Scenarios}
\begin{figure}[!tbp]
  \centering
  \begin{minipage}[b]{0.49\textwidth}
    \includegraphics[width=0.9\linewidth]{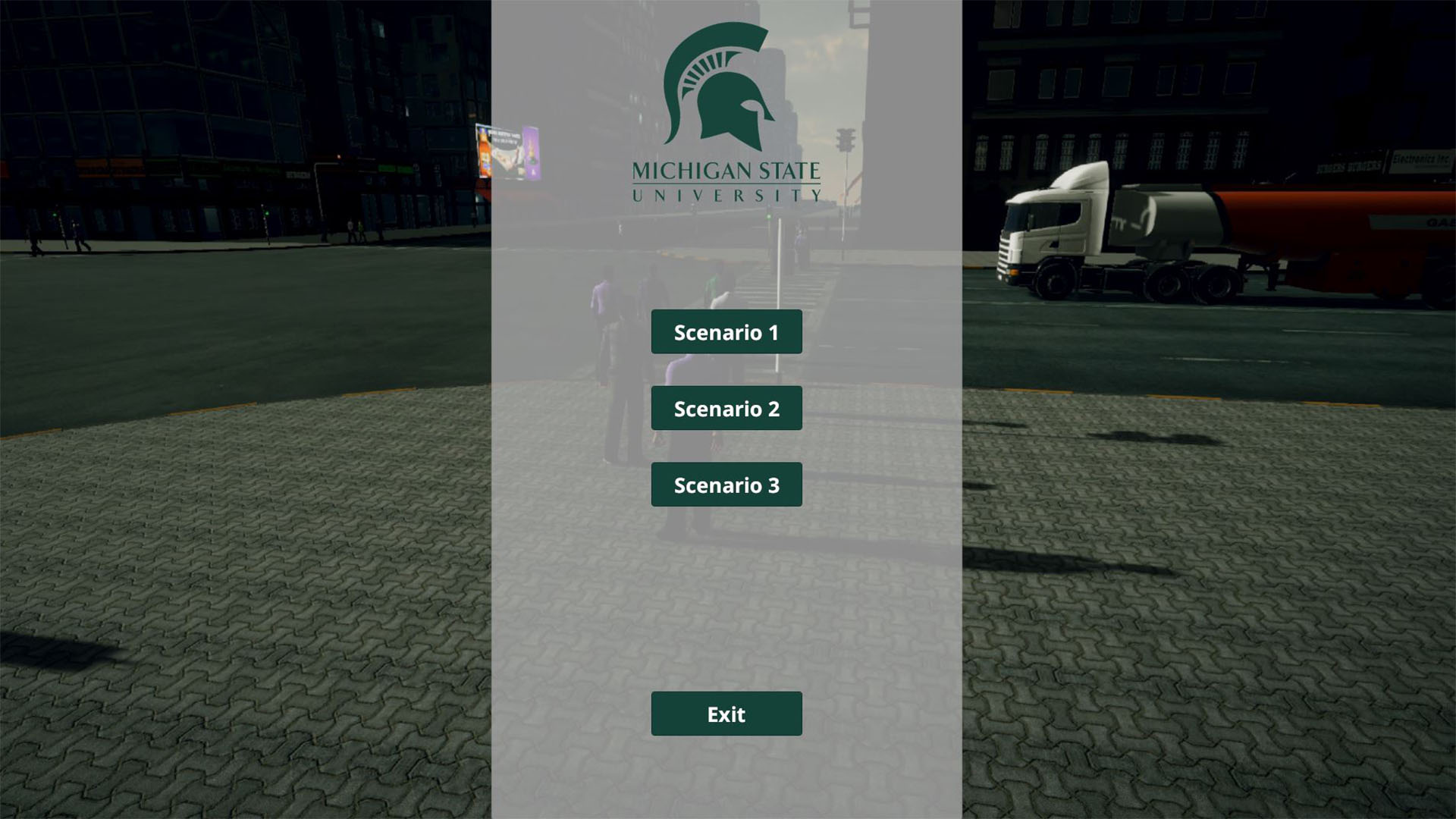}
    \caption{Main Menu helps the participants to switch between the various scenarios of the tool}
    \label{MainMenu}
  \end{minipage}
  \begin{minipage}[b]{0.49\textwidth}
    \includegraphics[width=0.9\linewidth]{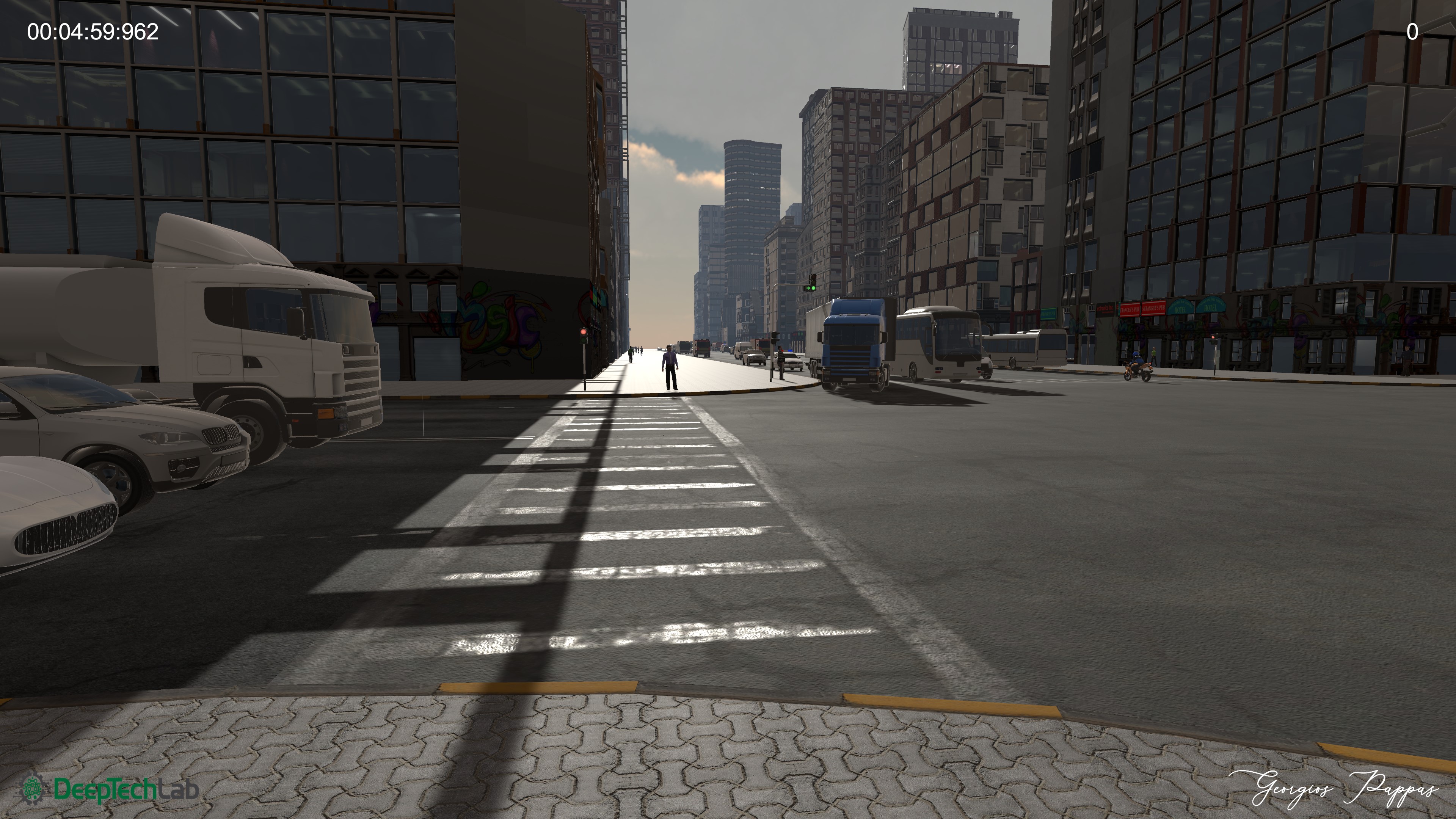}
    \caption{The urban area of the three scenarios features various buildings, vehicles, a skybox and a weather system}
    \label{UrbanArea}
  \end{minipage}
  \begin{minipage}[b]{0.49\textwidth}
    \includegraphics[width=0.9\linewidth]{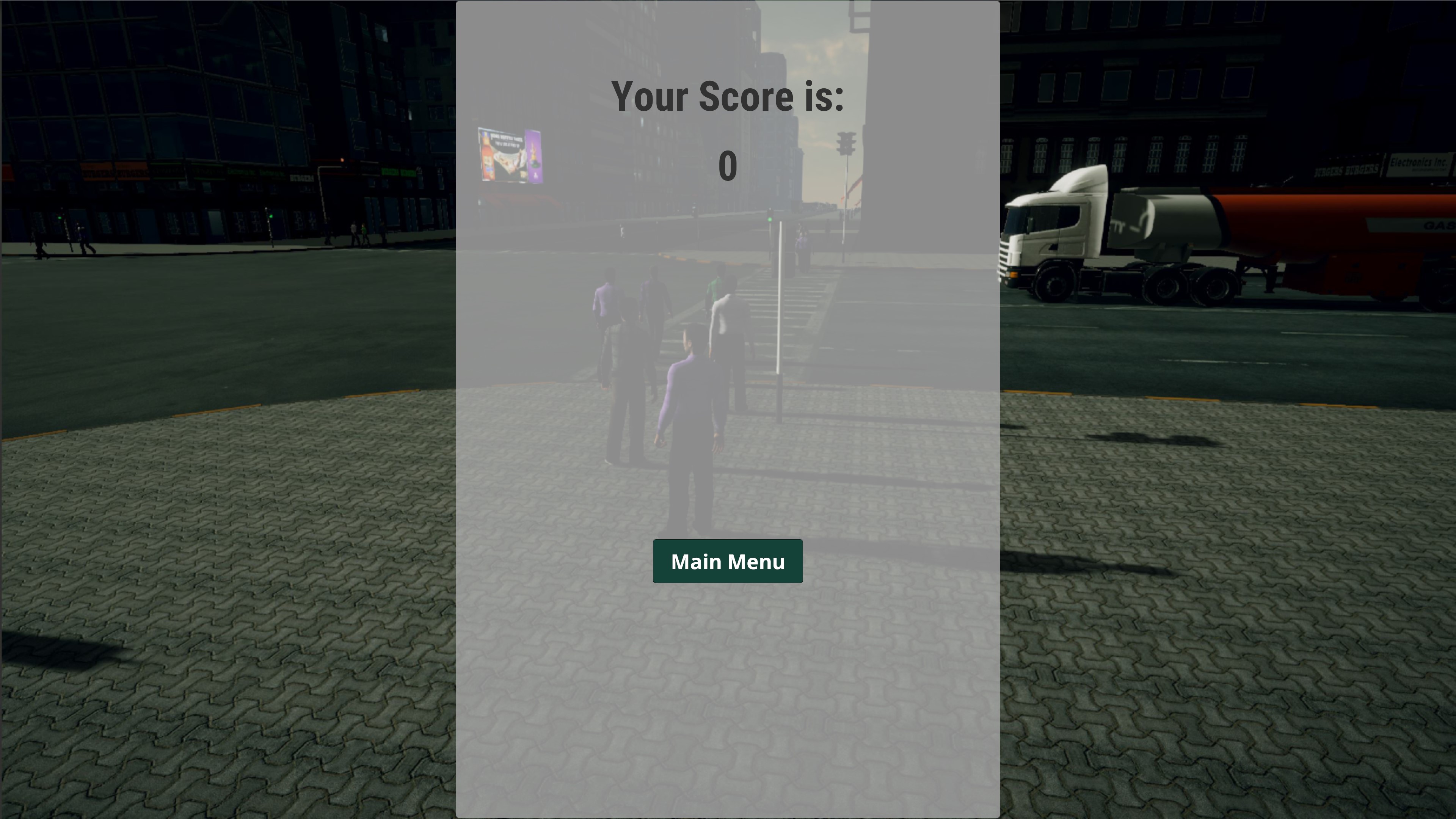}
    \caption{The Score / Game Over Scene shows the participants score based on their successful intersection crossings.}
    \label{GameOver}
  \end{minipage}
    \begin{minipage}[b]{0.49\textwidth}
    \includegraphics[width=0.9\linewidth]{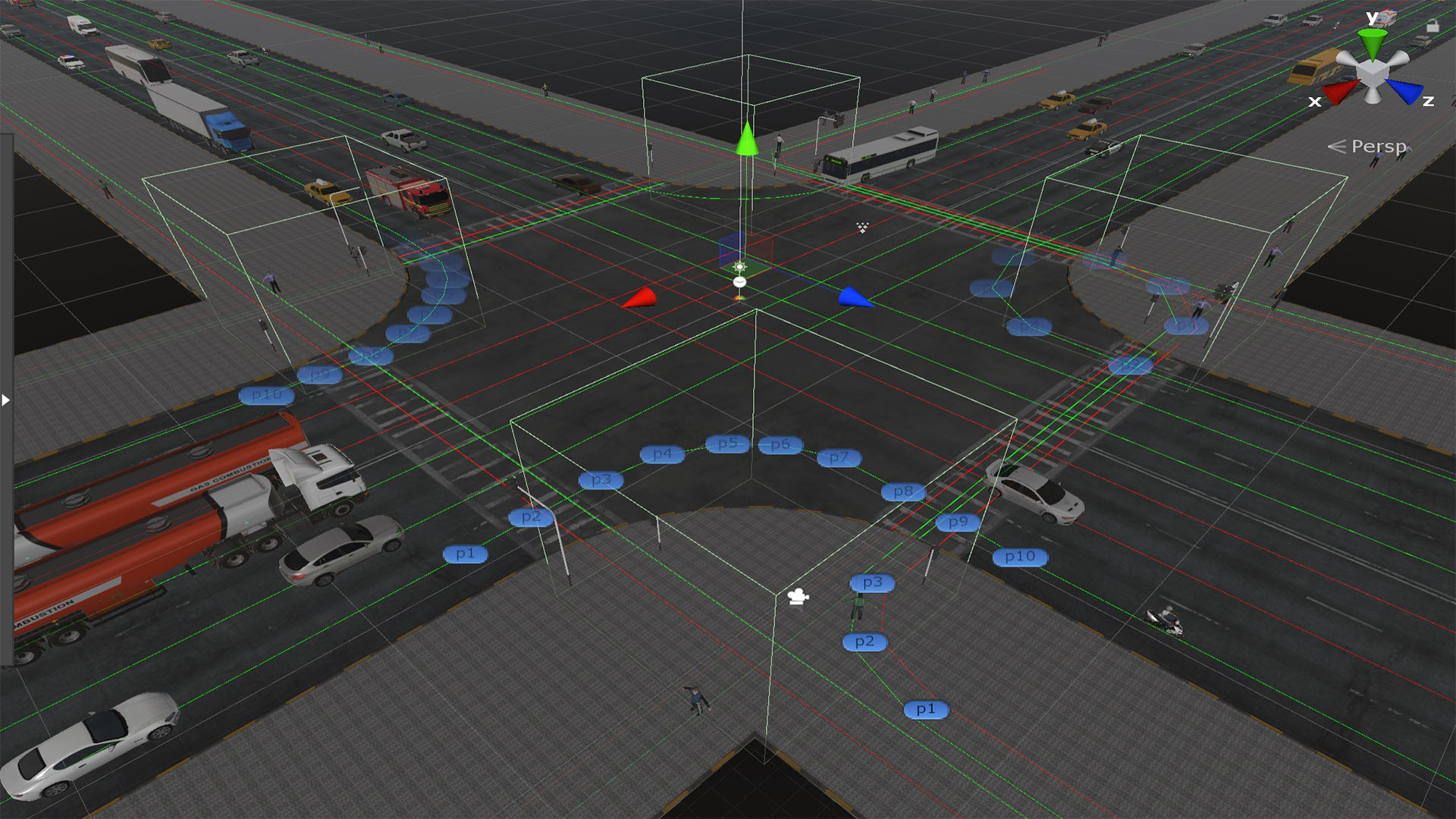}
    \caption{The invisible collider system is the main mechanic that the scoring system is built upon.}
    \label{ScoreColliders}
  \end{minipage}
\end{figure}

The tool is developed in Unity Game Engine. For its development, we used both free and purchased assets from Unity Asset Store. It is a first person perspective gamified application and it features a main menu (Figure \ref{MainMenu}), three different scenario scenes in an urban environment, and a game over scene (Figure \ref{GameOver}).

Once the game is loaded to main menu, the participant selects one of the three scenarios. These three scenarios are taking place in the same urban environment but the cars (virtual AVs and human driven) are having different behaviors. 

\begin{figure}
 \centering
 \includegraphics[width=0.6\linewidth]{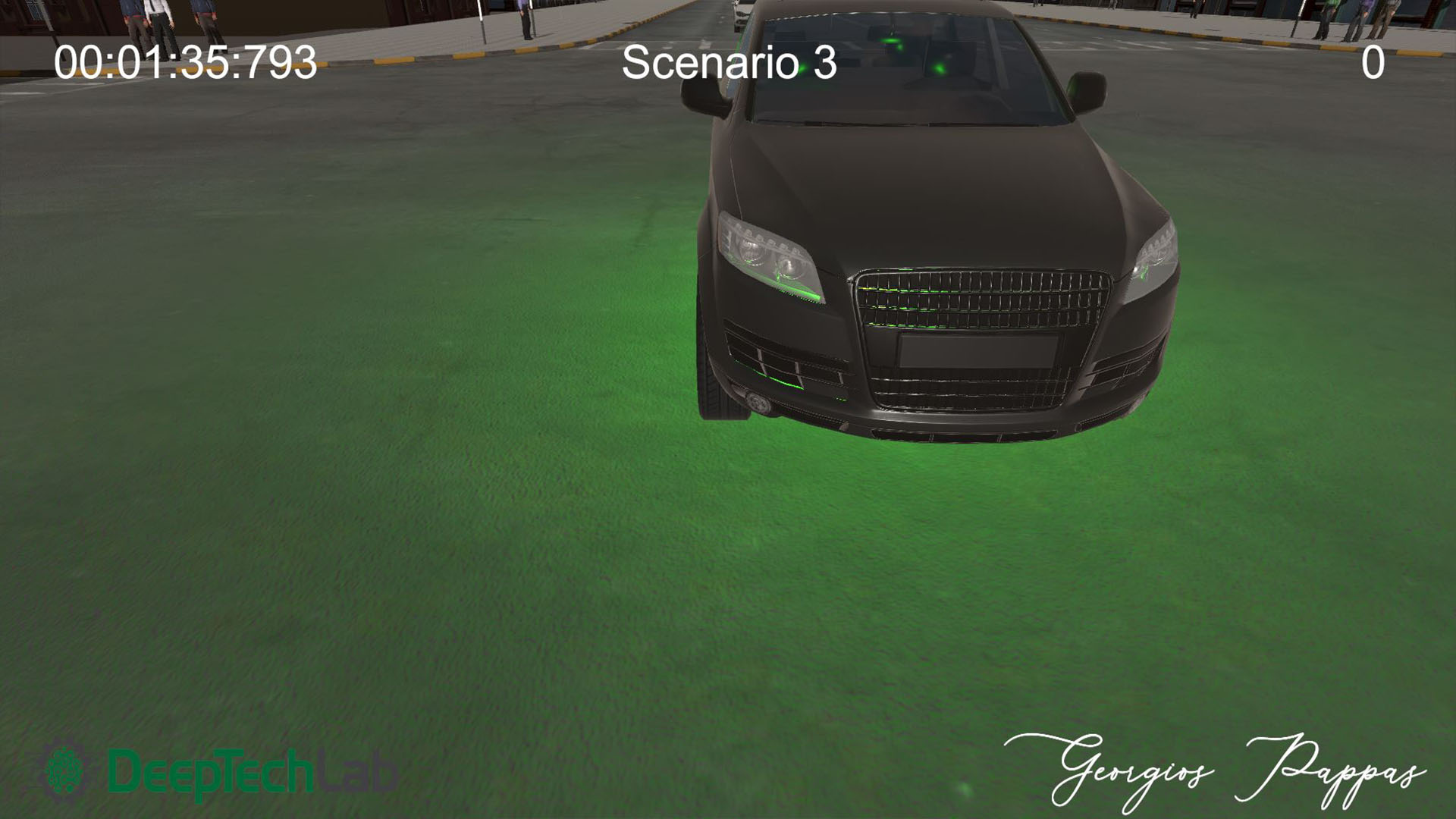} 
 \caption{In scenario 3, the AVs are identified by the green light underneath. The light appears when the pedestrian is 15 meters or less away from the vehicle.}
 \label{Scenario3}
\end{figure}

\begin{itemize}
    \item \textbf{Scenario 1:} All cars, human or AI operated, stop safely before hitting the user-pedestrian if physically possible. The possibility of being ``hit'' is low.
    \item \textbf{Scenario 2:} Some cars may stop before colliding with the user-pedestrian. Users do not know which if a given car (AV or human driven) will stop or not. The possibility of being ``hit'' is increased.
    \item \textbf{Scenario 3:} Some cars may stop before the user-pedestrian while others will not. In this case, the AVs are identified by a green light indication that appears under the vehicles. This indication starts when the pedestrian-AV distance is less than 15 meters. The AVs will always stop before hitting the pedestrian if physically possible (Figure \ref{Scenario3})
\end{itemize}

\subsubsection{Mechanics over various application builds}
The general idea of the training is that the users will try to cross intersections safely as virtual pedestrians while at the same time they develop awareness on self-driving vehicles. 

The first build of our application featured only the urban scene and the users could walk and cross intersections freely without any consequence even if they were 'virtually' hit by a vehicle. To make the simulation more engaging and fun ~\citep{4Keys2fun}, since the second version of the tool, there were some new implementations developed. Initially, we created a \textbf{scoring system}. Thus, for every successful crossing a user gets 100 points. To achieve this, we created a non-visible collider system (Figure \ref{ScoreColliders}) that works as trigger. Specifically, for the urban scene there are four colliders, one at each corner of the pavements. Once the user gets into it, a variable gets the name of the pavement and compare it with the last name it had in memory. If it is different (meaning that a user has crossed an intersection), it adds 100 points to the UI scoring system appearing at the top right corner of the screen.

Then, we also implemented a \textbf{timer} to limit the time of the training. Most importantly, since the second build of the application, ``careless'' users may result in losing at the game if they are being ``hit'' by a vehicle. This \textbf{penalization} implementation is accomplished by another collision system achieved between the Player component and the spawning vehicle. Once a user is ``hit'' or the time ends up, the simulation automatically transitions to the Game Over scene (Figure \ref{GameOver}) where the final score is shown. 

In the third and last major build, we implemented three scenarios (see Subsection \ref{Scenes-Scenarios}) with differentiation in vehicle behavior useful for data collection. All the scenarios are based on the same urban scene.

\section{Case Study - UX Experimentation}
\subsection{Step 1: Behavior in Reality vs no Consequence Virtuality}
\begin{figure}
 \centering
 \includegraphics[width=0.7\linewidth]{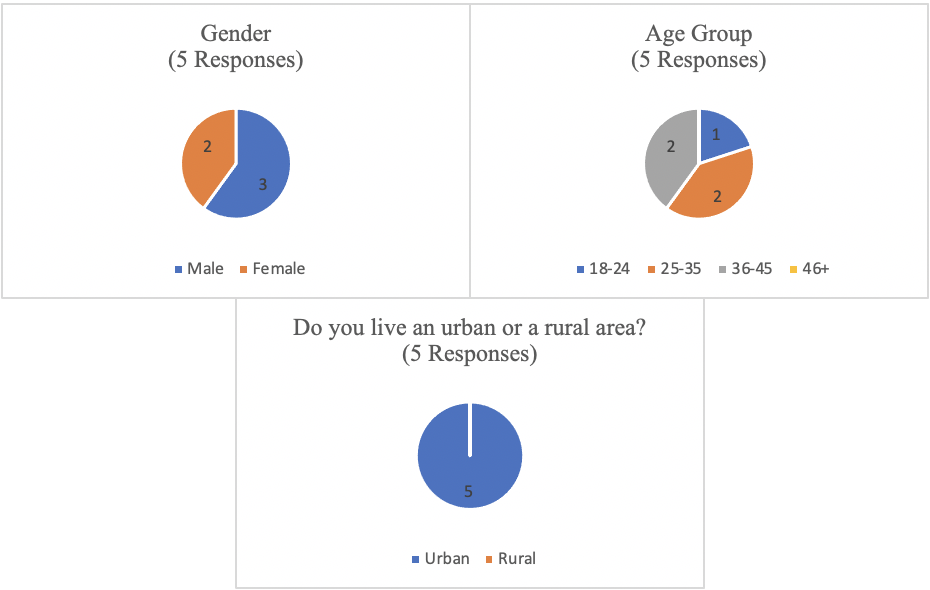} 
 \caption{Basic demographic information of the participants}
 \label{DemographicsFigureStep1}
\end{figure}

Based on the first build of our application, We conducted a first UX experiment. The goal of this experiment was to understand how people crossed an intersection in a simulation without knowing whether the vehicles in the road were automated vehicles or not, or even if they would stop if the user walked out in front of them. We conducted the experiment with five participants. They were between the ages of 18-45, represented mixed genders, and we purposely chose people who live in urban environments (Figure \ref{DemographicsFigureStep1}), as the first build had only this option available. Demographics for each participant can also be found at the Table \ref{tab:ParticipantDemographicsStep1}.

\begin{table}[]
    \centering
    \resizebox{0.5\linewidth}{!}{%
    \begin{tabular}{c | c c c}
    \hline
    \hline
    \textbf{Participant} & \textbf{Gender} & \textbf{Age Group} & \textbf{Area they live in} \\
    \hline
    \hline
    \textbf{Participant 1} & Male & 25-34 & Urban\\
    \hline
    \textbf{Participant 2} & Male & 35-44 & Urban\\
    \hline
    \textbf{Participant 3} & Female & 35-44 & Urban\\
    \hline
    \textbf{Participant 4} & Male & 18-24 & Urban\\
    \hline
    \textbf{Participant 5} & Female & 25-34 & Urban\\
    \hline
    \hline
    \end{tabular}
    }
    \caption{Basic demographic information of the participants}
    \label{tab:ParticipantDemographicsStep1}
\end{table}

\begin{figure}
 \centering
 \includegraphics[width=0.6\linewidth]{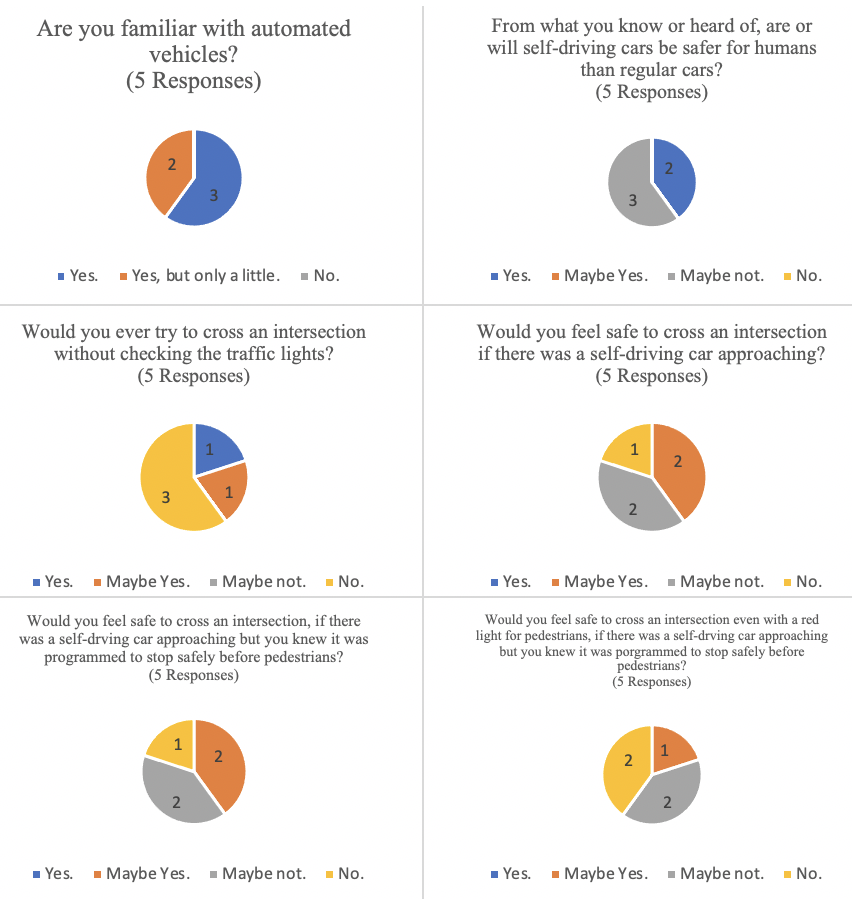} 
 \caption{Participant responses to questions relating to automated vehicles.}
 \label{Step1Survey}
\end{figure}

We began the experiment by asking each participant the same questions, regardless of demographic. These questions included their feelings about self-driving vehicles such as how familiar they are with self-driving vehicles, if they think self-driving vehicles will become a safer alternative to human-driven vehicles, and various questions about different scenarios of whether or not participants would cross the street in scenarios involving driving cars and safety. Responses to these questions are represented at Figure \ref{Step1Survey}. After participants finished the survey, they were given five minutes in the simulation to cross the intersections of our virtual city and experience the simulation, during which observers took note of statistics. After the five minutes were up, the participants were asked some follow-up questions, which included whether or not each participant found getting ``hit'' by the car ``fun'', what their opinions on the overall aesthetics were, and finally, just stating one thing about their overall experience with the tool.
The results of the experiment were tracked using the following rubrics:

\begin{itemize}
    \item \textbf{Rubric 1:} How many attempts of crossing intersections were made?
    \item \textbf{Rubric 2:} How many of the crossing were done “safely” (checked traffic lights etc.)?
    \item \textbf{Rubric 3:} Did the user try to cross just to check if they will be hit by a vehicle?
    \item \textbf{Rubric 4:} Did the user find the idea of ``being hit by a car'' in the game ``fun''?
    \item \textbf{Rubric 5:} How many times did they walk and how many did they run while crossing an intersection?
\end{itemize}

\begin{table}[]
    \centering
    \resizebox{0.6\linewidth}{!}{%
    \begin{tabular}{c | c c c c c}
    \hline
    \hline
    \textbf{Participant} & \textbf{Rubric 1} & \textbf{Rubric 2} & \textbf{Rubric 3} & \textbf{Rubric 4} & \textbf{Rubric 5}\\
    \hline
    \hline
    \textbf{Participant 1} & 10 & 9 & Yes & Yes & 8/2\\
    \hline
    \textbf{Participant 2} & 12 & 6 & Yes & Yes & 3/9\\
    \hline
    \textbf{Participant 3} & 10 & 2 & Yes & No & 5/5\\
    \hline
    \textbf{Participant 4} & 13 & 2 & Yes & Yes & 0/13\\
    \hline
    \textbf{Participant 5} & 11 & 11 & No & No & 7/4\\
    \hline
    \hline
    \end{tabular}
    }
    \caption{Simulation Observations/Question Responses}
    \label{tab:SimulationObservationsStep1}
\end{table}

This experiment taught us a lot, with one major finding from the results of the automated vehicle questions (Figure \ref{DemographicsFigureStep1}) being that the participants generally didn't trust automated vehicles as being safe in various scenarios. To go along with this, most of the participants wouldn't consider crossing an intersection without checking traffic lights and 80 percent of participants wouldn't cross the intersection if the traffic lights told them that they couldn't cross. This shows that most of the participants generally would not take risks in real life to increase efficiency to get to their destination due to the unsafe implications of their actions. However, despite these results from the preliminary survey, observations from their simulation experiences (Table \ref{tab:SimulationObservationsStep1}) show that four out of five participants still crossed the street in an unsafe manner in the simulation despite their tendencies to do the exact opposite in real life.

This led to our conclusion that people act differently in virtuality than in reality, and the actions of the participants as well as their post-questionnaire results further prove this. Of the four participants who attempted to walk in front of a car, three of them found it ``fun'' to get ``hit'' by the car, with one participant citing ``being able to push the cars'' as one of their favorite features. Based on the participants' pre-survey results, they would never consider actions such as those described above in real life as they would be putting themselves at risk. Participants stuck to this ideology early on the experiment. However, they soon realized that their actions didn't have consequences, so most of the participants decided to be riskier with their intersection crossing and soon experienced being ``hit'' by a car. This needed to be well improved upon the next versions of the tool to improve realism. 

\subsection{Step 2: Behavior in Reality vs Consequence Virtuality}
Once the second version of the tool was developed, we conducted another UX experiment with four participants this time. the demographics along with more information on the participants are shown at the table \ref{tab:ParticipantDemographicsStep2}.

\begin{table}[]
    \centering
    \resizebox{0.8\linewidth}{!}{%
    \begin{tabular}{c | c c c c c}
    \hline
    \hline
    \textbf{Participant} & \textbf{Gender} & \textbf{Age Group} & \textbf{Experience with} & \textbf{Experience with} & \textbf{Experience with} \\
     &  &  & AVs (usage) & AVs (General Knowledge) & games (especially \\
     &  &  &  &  & vehicle-related like GTA) \\
    \hline
    \hline
    \textbf{Participant 1} & Male & 18-24 & \FiveStar \FiveStarOpen \FiveStarOpen \FiveStarOpen \FiveStarOpen & \FiveStar \FiveStar \FiveStarOpen \FiveStarOpen \FiveStarOpen & \FiveStar \FiveStar \FiveStar \FiveStar \FiveStarOpen\\
    \hline
    \textbf{Participant 2} & Male & 18-24 & \FiveStar \FiveStarOpen \FiveStarOpen \FiveStarOpen \FiveStarOpen & \FiveStar \FiveStar \FiveStarOpen \FiveStarOpen \FiveStarOpen & \FiveStar \FiveStar \FiveStar \FiveStar \FiveStar\\
    \hline
    \textbf{Participant 3} & Male & 35-44 & \FiveStar \FiveStarOpen \FiveStarOpen \FiveStarOpen \FiveStarOpen & \FiveStar \FiveStar \FiveStarOpen \FiveStarOpen \FiveStarOpen & \FiveStar \FiveStar \FiveStar \FiveStar \FiveStar\\
    \hline
    \textbf{Participant 4} & Female & 18-24 & \FiveStar \FiveStar \FiveStar \FiveStar \FiveStar & \FiveStar \FiveStar \FiveStar \FiveStar \FiveStar & \FiveStar \FiveStarOpen \FiveStarOpen \FiveStarOpen \FiveStarOpen\\
    \hline
    \hline
    \end{tabular}
    }
    \caption{Demographic data and qualitative experience information as reported by study participants}
    \label{tab:ParticipantDemographicsStep2}
\end{table}

This tool version offered a timer, a score mechanism for successful intersection crossings (100 points for each) and most importantly, a penalization system resulting in ending the game when a participant was virtually ``hit'' by a vehicle. 

The users playtested the tool for two minutes each. Due to COVID-19 situation, the tool was sent and the observation took place remotely. One of the main constraints identified was that not all users had high-end computers. Thus, the experience was quite different in terms of realism when played on ultra vs medium or even vs low graphics.

The most important insight that we got by observing the users was that their behavior towards vehicles changed drastically. Although this behavior was not exactly natural, users started to avoid cars and tried either to follow the pedestrian traffic lights as in real life or even to cross more strategically. But this was a big step getting a more realistic behavior in virtuality. 

An additional insight we got by observing the users, is that some of them tended to explore the overall map instead of just completing their crossing objective. This behavior is known as an ``explorer'' type of player based on Bartle's Taxonomy~\citep{Bartle1996}~\citep{MarvinOliverSchneiderErikaTiemiUeharaMoriyaJoaoCarlosNeto2016}. Although this exploring action was really useful for playtesting purposes and for identifying bugs, it also led us to conclude that training needed to be supervised rather than allowing participants to freely use the application as a game.

\subsection{Step 3: Testing AV Interaction \& Behavioral Intentions}

Following the first and second user experience tests, a final experiment was conducted with nine participants to better understand user interactions with autonomous vehicles and predictability. Final sample for this step included seven valid responses for analysis.

Including additional minor updates from UX testing, the tool was equipped with the same timer and scoring mechanism for successful intersection crossings and penalizations for being ``hit''. Participants were recruited through convenience sampling methods and were again observed remotely due to COVID-19 protocols. 

Differing slightly from the prior format, each user interacted with three scenarios for two minutes each. The three scenarios all utilized the same controls and depicted the same environment, but were identifiable as follows: the first scenario included vehicles that always stopped for pedestrians, the second scenario included some vehicles that would stop for pedestrians and some that would not, and the third scenario included some that would stop and others that would not, but autonomous vehicles would indicate their presence with a bright green light when they were within a 15 meter radius from the user. 

In order to test the behavior and perceptions across each of these three scenarios, user behaviors were observed and each user was provided with a Qualtrics survey to record their responses. Baseline questions regarding real world scenarios, pedestrian behaviors, and technological and autonomous vehicle familiarity were obtained prior to beginning the pedestrian experience, and then a series of questions were answered following exposure to each of the individual scenarios. Questions following the scenarios focused on the user perceptions of the autonomous vehicle tool and the specific actions taken during their scenario experience. Following the final scenario, additional questions were answered regarding general user experience, usability ~\citep{brooke1996sus}, realism ~\citep{mcgloin2013video}, and demographics.

Pre-tests were used to evaluate current behaviors and familiarity with the present topics. This data concluded that the majority of participants would never try to cross an intersection without checking traffic lights in the real world (71\%). Additionally, when asked if they would feel safe crossing an intersection and they knew there was a self-driving car approaching, the majority of participants responded ``maybe not'', regardless of the green (OK to cross) or red signal (do not cross) to cross. Additional data concluded that participants were somewhat unlikely to trust AVs to stop for an object or person in the road (M=4.7, Likert scale 1-10) and neutrally to slightly less familiar with AVs to begin with (M=4, Likert scale 1-10).

The data from the post-tests of these three scenarios led to some valuable insight. When asked about the noticeable presence of self-driving cars in the simulation, responses to scenario two (53\% noticed) and scenario three (69\% noticed) were as expected with an increasing awareness of self-driving vehicles. It is to be noted that some participants (37\%) reported noticing self-driving cars in the first scenario when they were not present. Furthermore, when asked about the clear intentions of self-driving vehicles in the simulation, participants in scenario two were mostly unsure (57\%) of intentions while a majority of those in scenario three reported that intentions were clear (57\%). When asked about their intentions to follow traffic signals in the scenario, the likelihood of following signals progressively decreased from being neutral to moderately unlikely as participants encountered additional scenarios with an increased presence of self-driving vehicles. As for trusting an AV to stop in the simulation, participants became less trusting of the self-driving vehicles as their presence increased across the scenarios (scenario 1: slightly likely, scenario 2: slightly unlikely, scenario 3: moderately unlikely). 

When it came to specific user behaviors in the simulation, scenario one had 71\% of participants running through the simulation and then scenarios two and three with 86\% of participants running through the simulation. It is to be noted that participants had the option to run, walk, or a mix of both. Additionally, participants reported becoming increasingly cautious at the onset of scenario three, as opposed to careless in their behaviors. In scenarios one and two 86\% and 71\% (respectively) of participants reported being careless, while only 43\% reported carelessness in scenario three. 

Broadly, general user experience data also confirmed that this simulation was easy to use, realistic, and not too complex. Some qualitative responses suggested additional control options for moving around the environment and additional weather scenarios, so that they could experience and perceive self-driving cars within differing environments. As for preferences on a consumer level, participants preferred that self-driving cars signal their presence to the public through audio alerts and/or lights or stickers that are more visible to pedestrians (i.e. on tops or sides of vehicles).

Finally, the demographic makeup of this sample was 100\% Caucasian, 71\% male, ages between 18 and 34, and a majority of participants (57\%) self-reporting that their primary residence was in a rural area. This sample size (N=7) and demographic makeup is a limitation of this study that does not allow for capturing the perceptions of a larger, more diverse sample of the population. 

\section{Conclusions}
The present research provided some valuable information on the user perspective on AVs. Initially, the users' survey gave us a first indication that people are not so ready to trust autonomous vehicles being on the streets. This distrust does not only include being passengers in such vehicles but also facing them as pedestrians. Further to that, the same group of people, even though they would hesitate crossing an intersection when a self-driving car approached in real life, their behavior was different in the virtual world using the first version of the tool. Specifically, since they had no consequences when they were virtually ``hit'' by a vehicle they tended to be careless or even abuse cars when they stopped before them. Using the next version of the tool, where a reward and a penalization system were implemented, users started behaving closer to reality. Their approach was more strategic but surely they were more careful in virtuality since being ``hit'' by a vehicle (self-driven or not) resulted in a game over.

Finally, the final testing version of the tool provided us with both technical and human factor insights. Firstly, it was further confirmed that that the tool itself is usable, realistic, and not too complex for simulation use. Secondly, it was highlighted that users become more cautious when AVs are present and mixed scenarios of AV/Non-AV presence allow for greater uncertainty among participants, when it comes to AV intentions. One contradictory point is that although users are increasingly less likely to obey traffic signals, they are also increasingly less likely to trust AVs. This could speak to the gamification aspect of the simulation, in conjunction with the overwhelming behavior to run and not walk throughout the simulation. Additionally, users reported the presence of AVs in scenario one, when they were not present. This could be a matter of clarity regarding the scope of the simulation, or an increased uncertainty of AV signaling an intentions. Further research is needed regarding these points and what impact that has on the behavior translation to the real world. 

Differences in the whole experience were noted where users tested the tool in medium or low graphics. The performance of their computers played a significant role in how well they were immersed in the application.

Overall, users seemed to enjoy the tool especially, the experienced gamers who were at the same time inexperienced with AVs. They have found the tool easy-to-use since the controls were quite similar to popular games.

\section{Future Work}
The present work resulted in developing a first person gamified research tool about pedestrian-autonomous vehicle interaction. In the future, we intend to create a VR version of the tool for users to have a fully immersive experience. Moreover, we intend to have additional intersections and probably more levels in order for users to have more tasks and they would not have to repeat on the same environment. Regarding the AI pedestrian models we will add more or customize existing ones in order to make the tool even more diverse ~\citep{Kafai2016}.Upon completing all the development parts, we also plan to release the tool freely for further use and research by the community.

As for opportunities for human factors research, this tool provides a unique option for studying public perceptions and behavioral intentions surrounding AVs. Additional studies could conduct larger, more diverse scale testing, comparative studies of simulation, VR, and reality, additional scenario studies surrounding urban vs rural environments and varying weather conditions, as well as longitudinal human factor studies surrounding trust, acceptance, perceived risk, and interaction preferences with consumer implications.

\bibliographystyle{unsrtnat}
\bibliography{references}
\end{document}